\documentclass[aps,groupedaddress,amsmath,amssymb,reprint]{revtex4-1}

\usepackage{graphicx}
\usepackage{bm}

\begin{document}


\title{Ambipolar High Mobility Hexagonal Transistors on Hydrogen-Terminated Silicon (111) Surfaces} 



\author{Binhui Hu}
\email[To whom correspondence should be addressed. Electronic mail:
] {hubh@umd.edu}
\author{Mohamad M. Yazdanpanah}
\author{Joyce E. Coppock}

\author{B. E. Kane}
\affiliation{Laboratory for Physical Sciences, University of
Maryland, College Park, Maryland 20740} \affiliation{Joint Quantum
Institute, University of Maryland, College Park, Maryland 20742}
\date{\today}



\date{\today}

\begin{abstract}
We have fabricated ambipolar transistors on chemically prepared
hydrogen-terminated Si(111) surfaces, in which a two-dimensional
electron system (2DES) or a two-dimensional hole system (2DHS) can
be populated in the same conduction channel by changing the gate
voltage of a global gate applied through a vacuum gap. Depending on
the gate bias, ion implanted n$^+$ and p$^+$ regions function either
as Ohmic contacts or as in-plane gates, which laterally confine the
carriers induced by the global gate. On one device, electron and
hole densities of up to $7.8\times10^{11}$ cm$^{-2}$ and
$7.6\times10^{11}$ cm$^{-2}$ respectively are obtained. The peak
electron mobility is $1.76\times10^5$ cm$^{2}$/Vs, and the peak hole
mobility is $9.1\times10^3$ cm$^{2}$/Vs at 300 mK; the ratio of
about 20 is mainly due to the very different valley degeneracies
(6:1) of electrons and holes on the Si(111) surface. On another
device, the peak electron mobility of $2.2\times10^5$ cm$^{2}$/Vs is
reached at 300 mK. These devices are hexagonal in order to
investigate the underlying symmetry of the 2DESs, which have a
sixfold valley degeneracy at zero magnetic field. Three
magnetoresistance measurements with threefold rotational symmetry
are used to determine the symmetry of the 2DESs at different
magnetic field. At filling factor $1 < \nu < 2$, the observed
anisotropy can be explained by a single valley pair occupancy of
composite fermions (CFs). Qualitatively the CFs preserve the valley
anisotropy, in addition to the twofold valley degeneracy. At
magnetic field up to 35 T, the 2/3 fractional quantum Hall state is
observed with a well developed hall plateau; at $\nu<2/3$, the three
magnetoresistances show a large anisotropy (50:1). We also show that
device degradation is not a serious issue for our measurements, if
the device is kept in vacuum or a nitrogen gas environment and its
time in air is minimized.
\end{abstract}


\maketitle 

\section{INTRODUCTION}
Investigations of two-dimensional systems (2DSs) on Si surfaces
began with metal-oxide-semiconductor field-effect transistors
(MOSFETs) \cite{Ando82RMP, lakhani75, Neugebauer75, KlitzSi11074,
KlitzSi11174}. However, the amorphous SiO$_2$/Si interface severely
limits the quality of the 2DSs. The development \cite{Eng05APL,
Eng07PRL, McFarland09PRB, Hu10APL, Kott14PRB} of hydrogen-terminated
silicon (111) [H-Si(111)] vacuum field effect transistors (FETs) is
based on a simple but extraordinary fact: a wet chemical treatment
of a Si(111) surface with an ammonium fluoride (NH$_4$F) solution
can produce an ideal H-passivated Si surface, which is both
atomically flat and possesses a very low surface state density
($\leq 10^{10}$ cm$^{-2}$) \cite{Higashi90APL, Jakob91JCP,
Weinberger85JVST, Yablonovitch1986}, leading to much higher carrier
mobilities. In recent years, both a high quality two-dimensional
electron system (2DES) with electron mobility of 325,000 cm$^2$/Vs
\cite{Kott14PRB} and a high quality two-dimensional hole system
(2DHS) with hole mobility of 10,000 cm$^2$/Vs \cite{Hu10APL} have
been realized in this system. They are comparable to the best
Si/SiGe heterostructures \cite{Friedrich97SST, Lai04PRL, Lu12PRB,
whall94, basaran94}, which are limited to the (100) oriented
surfaces because of much higher threading dislocation densities on
the other surface orientations \cite{Lee06, Gatti14}.

In this work we have fabricated ambipolar hexagonal devices, which
can switch between a 2DES and a 2DHS in the same device by changing
a gate voltage. Compared to other ambipolar devices \cite{Chen12APL,
Croxall13APL, Novo05, Tian10, Das14, Betz14, Mueller15}, the 2DES on
the Si(111) surface has a sixfold valley degeneracy, while the
valley degeneracy of holes is one. The ambipolar devices provide a
direct comparison between electrons and holes with very different
valley degeneracies in the same conduction channel, which allows a
new perspective to the fundamental research, such as the properties
of 2D metals \cite{Hu15PRL}. In addition, the hexagonal devices
enable the exploration of the sixfold valley-degenerate electron
system.

In the H-Si(111) vacuum FET device, a global Si/SiO$_2$ gate piece
is used to induce a 2DS, which is further confined into a
well-defined 2D region by PN junction isolation and trench
isolation. Similarly, future nanoelectronic devices, like quantum
point contacts (QPCs) and quantum dots (QDs), may be realized on the
Si(111) surfaces using the in-plane depleting gates based on PN
junctions. Moreover, because of the sixfold valley degeneracy of the
2DES, the valley degree of freedom can provide extra resources for
practical applications in the same way as spintronic devices utilize
the spin degree of freedom. This will open up new opportunities for
another class of devices --- valleytronic devices
\cite{zeng12Nature, isberg13Nature, renard15Nature}.

The remainder of this paper is organized as follows. In
Sec.~\ref{sec:fab}, we describe the device structure and operation
principles, followed by detailed fabrication processes. In
Sec.~\ref{sec:characterization}, we characterize the devices to make
sure that they work as intended, including surface topography
measurement, a gate leakage check, PN junction isolation and leakage
tests, and contact resistance measurements. In
Sec.~\ref{sec:transport}, we discuss the transport measurement
results on these devices. We determined the 2D carrier mobilities in
the experimentally accessible range of densities, which show that
electron and hole mobilities are very different. Magnetotransport
measurements show rich phenomena, including Shubnikov-de Haas (SdH)
oscillations, integer quantum Hall effect (IQHE), fractional quantum
Hall effects (FQHE) and large anisotropy in three magnetoresistance
measurements at $\nu<2/3$. In Sec.~\ref{sec:degradation}, we discuss
the device degradation with time and with exposure to ambient air,
including the mobility deterioration and the increase of contact
resistance. We conclude in Sec.~\ref{sec:conclusion} with a
discussion of further improvements and future directions.

\section{\label{sec:fab}DEVICE STRUCTURE AND FABRICATION}

 \begin{figure}
 \includegraphics{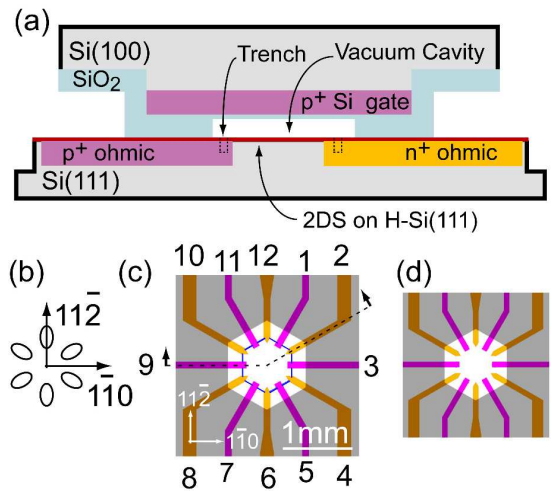}%
 \caption{\label{fig:schematics}(a) Schematic cross section of an
 ambipolar vacuum FET (not to scale), with a SiO$_2$/Si(100) remote
 gate piece and a H-Si(111) piece contact bonded in a vacuum. The
 remote gate piece is used to induce 2D carriers on the H-Si(111)
 surface by applying a gate voltage on a p$^+$ gate layer through an
 encapsulated vacuum cavity. It also has a fully covered SiO$_2$ layer
 with a thin layer of SiO$_2$ ($\sim$ 30 nm) left in the cavity.
 (b) Six degenerated valleys of the 2DES with crystallographic orientations. (c) (d) In
 the H-Si(111) piece, n$^+$ (p$^+$) Ohmic contacts are used to access a
 2DES (2DHS) at the center, with p$^+$ (n$^+$) regions acting as lateral
 confinement to restrict current flow between contacts to the hexagonal center 2DS. Two different
 configurations are investigated. (c) One has trench isolation,
 and (d) the other doesn't have trench isolation. The dash lines
 in (c) show the cross section depicted in (a).}
 \end{figure}

\begin{figure}
 \includegraphics{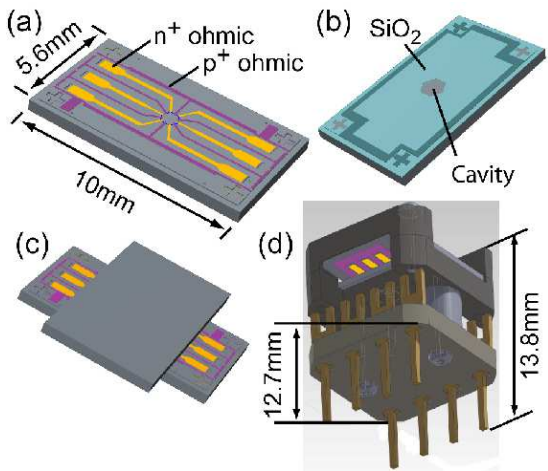}%
 \caption{\label{fig:realdevice}(a) A H-Si(111) piece with six n$^+$
 contacts (yellow) and six p$^+$ contacts (purple). (b) A Si/SiO$_2$ remote
 gate piece which is fully covered by SiO$_2$, except the two grey crosses
 where the gate contacts are made. There is a $\sim$ 30 nm oxide at the bottom
 of the hexagonal cavity. (c) 3D view of a bonded device. (d) A bonded
 device placed in a customized holder and installed on an 8-pin
 dual in-line package (DIP) header without wiring.}
 \end{figure}

The vacuum FET device is fabricated from two pieces. One is a
SiO$_2$/Si(100) remote gate piece, and the other is a H-terminated
high purity Si(111) piece, as shown in Fig.\
\ref{fig:schematics}(a). Just like a MOSFET, when a positive
(negative) gate voltage is applied between the remote gate and a
n$^+$ (p$^+$) source Ohmic contact through an encapsulated vacuum
cavity, a 2DES (2DHS) will be induced on the H-Si(111) surface.
Since the 2DES has a sixfold valley degeneracy [Fig.\
\ref{fig:schematics}(b)], the hexagonal device is designed to
investigate its underlying symmetry, with six n$^+$ Ohmic contacts
placed along the major axes of the constant energy ellipses on the
Si(111) surface, labeled as 2,4,$\dots$, 12. It also has six p$^+$
Ohmic contacts placed between the n$^+$ contacts, labeled as
1,3,$\ldots$,11. The remote gate piece with a hexagonal cavity
[Fig.\ \ref{fig:realdevice} (b)] is contact bonded on the H-Si(111)
piece in a vacuum chamber. When a positive (negative) gate voltage
is applied, the n$^+$ (p$^+$) Ohmic contacts are used to access the
2DES (2DHS) at the center, and the p$^+$ (n$^+$) contacts act as
lateral confinement. Figure \ref{fig:schematics} (b)-(d) show the
arrangement of the n$^+$ and p$^+$ Ohmic contacts relative to
crystallographic directions.

Two different types of devices are investigated. One has shallow
trench isolation (STI) to confine the center 2DS [the center blue
hexagon in Fig.\ \ref{fig:schematics}(c)]; the other doesn't have
the STI, and the center 2DS is defined by the hexagonal cavity
[Fig.\ \ref{fig:schematics}(d)]. There is typically a $\sim$ 100
$\mu$m misalignment between the remote gate piece and the H-Si(111)
piece when they are contact bonded. Because the distance between the
trenches and the edges of the hexagonal cavity is more than 200
$\mu$m, and the trenches are etched on the Si(111) piece, the
misalignment doesn't change the center 2DS in the device with the
STI; the misalignment can affect the center 2DS in the device
without the STI when the n$^+$ and p$^+$ contacts are not concentric
and aligned to the hexagonal cavity. However, the measurement
results show that there is no significant difference between these
two types of devices, because the relative placement of the contacts
is fixed, and the bonding alignment is not critical in defining the
device with the edges of the cavity far outside the contact
perimeters. We have fabricated six devices (194-199) with peak
electron mobilities of $2.2\times10^5$ cm$^{2}$/Vs, $5.6\times10^4$
cm$^{2}$/Vs, $2.0\times10^5$ cm$^{2}$/Vs, $5.2\times10^4$
cm$^{2}$/Vs, $1.76\times10^5$ cm$^{2}$/Vs and $5.8\times10^4$
cm$^{2}$/Vs at 300 mK  respectively, in which four (194-197) of them
have the STI, and the other two (198,199) do not have it. Most data
presented here are from two devices, sample 194 (with the STI) and
sample 198 (without the STI).

\subsection{Gate piece}

The remote gate piece is fabricated from a Si(100) wafer (float
zone, Si:B, resistivity $>10,000 \Omega\cdot$cm). First, a p$^+$
gate conducting layer is formed by ion implantation with
$2.4\times10^{15}$ cm$^{-2}$, 15 keV boron ions, as shown in Fig. \
\ref{fig:schematics}(a). Then a 2 $\mu$m deep Si mesa is dry etched
around the edges of each die using reactive ion etching (RIE). Next,
a $\sim$ 350 nm thick SiO$_2$ layer is thermally grown on the top:
dry oxidation at 1050 $^\circ$C for 15 min, followed by wet
oxidation at 1050 $^\circ$C for 27 min and another dry oxidation at
1050 $^\circ$C for 10 min. The dry-wet-dry cycle is used to improve
the gate oxide quality, and also activates the implanted dopants.
After that, a cavity with a $\sim$ 30 nm thick oxide layer left
behind is formed by dry etching and wet etching. In order to make
sure the oxide thickness is reduced to $\sim$ 30 nm in the cavity, a
test wafer and the device wafer are loaded into a RIE machine and
dry etched at the same time. Both have the same $\sim$ 350 nm thick
oxide layer, and the oxide layer is dry etched by $\sim$ 270 nm.
Then the depth of the cavity in the test wafer is measured using a
profilometer after removing photoresist. The time of the wet etching
to leave the $\sim$ 30 nm thick oxide layer is calculated based on
the oxide thickness, the depth of the cavity and the wet etching
rate. The fully SiO$_2$ encapsulated gate piece with the oxide
reduced cavity drastically reduces the likelihood of gate leakage,
compared to the gate piece where the doped regions were exposed. We
have tested remote gate pieces with either $\sim$ 350 nm or $\sim$
200 nm thick thermal oxide, and both of them work well. Finally, the
wafer is diced into $5.6\times10$ mm$^2$ remote gate pieces, as
shown in Fig.\ \ref{fig:realdevice} (b).

\subsection{Si(111) piece}

The H-Si(111) piece is also $5.6\times10$ mm$^2$ made from a p$^-$
high purity Si(111) wafer (float zone, resistivity $>10,000
\Omega\cdot$cm ). First, a 30 nm thick sacrificial SiO$_2$ layer is
thermally grown on the top of the Si(111) substrate by dry oxidation
at 950 $^\circ$C for 30 min in a CMOS grade furnace to reduce the
channeling effect during ion implantation and protect the clean
Si(111) surface. Next, the six n$^+$ contact regions are patterned
by photolithography with the primary flat of the Si(111) wafer
(oriented to [$11\bar{2}$] \cite{[{SEMI International standards,
SEMI M1-0298}][{}]Sistandard}) aligned ($< 5 ^\circ$) along the long
side of the dies. Then the wafer is ion implanted with
$4.5\times10^{14}$ cm$^{-2}$ phosphorus ions at 50 keV. Alignment
marks are formed at the same time. Similarly, the six p$^+$ contact
regions are defined by photolithography and ion implantation with
$9\times10^{14}$cm$^{-2}$ boron ions at an energy of 15 keV. The ion
implantation parameters are based on previous works \cite{Eng05APL,
Hu10APL} and simulation results from SUPREM-IV process simulator
\cite{[{SUPREM-IV computer code was developed by R. W. Dutton and J.
D. Plummer at the Integrated Circuits Laboratory, Stanford
University}][{}]suprem4}. The parameters are selected so that after
thermal annealing, the peak doping concentration is located near the
Si(111) surface, and about one order of magnitude higher than that
of the three-dimensional (3D) metal-insulate transition (MIT), as
shown in Fig.\ \ref{fig:Dopingprofile}. (The 3D MIT occurs at doping
concentration of $3.74\times 10^{18}$ cm$^{-3}$ for phosphorus
\cite{Rosenbaum80} and $3.95\times 10^{18}$ cm$^{-3}$ for boron
\cite{Dai91}.) After each ion implantation, the photoresist is
removed by acetone cleaning and piranha cleaning. First the wafer is
immersed in boiling acetone for 30 min to remove most of the
photoresist. Then the wafer is rinsed with isopropyl alcohol (IPA)
and de-ionized (DI) water. Next, the wafer is immersed in a piranha
solution (4 H$_2$SO$_4$(98$\%$) : 1 H$_2$0$_2$ (30$\%$)) at 100
$^\circ$C for 30 min to remove the residual photoresist. Right
before rapid thermal annealing (RTA), the wafer is RCA-1 cleaned for
5 min with a H$_2$O$_2$/NH$_4$OH/H$_2$O solution, then thoroughly
rinsed in DI water, and spin-dried. The wafer is annealed in a CMOS
grade rapid thermal annealer at 950 $^\circ$C for 60 sec. Alignment
marks from ion implantations are visible before the RTA process, but
invisible after it. RCA-1 clean etches the ion implanted SiO$_2$
regions before the RTA process. They serve as alignment marks to do
mesa etching, but the contrast is very low. It is helpful to define
global alignment marks before the RTA process using CMOS compatible
processes. Next, if STI is desired, the 200 nm deep and 10 $\mu$m
wide shallow trenches are formed by dry etching. After that, a 2
$\mu$m deep Si mesa is dry etched around the edges of each die. This
mesa prevents particles from accumulating on the surface while
handling the substrate and ensures a clean edge for bonding
\cite{Eng05APL}. It also defines each die for dicing. Finally, the
wafer is diced into individual Si(111) pieces, as shown in Fig.\
\ref{fig:realdevice}(a).

\subsection{Bonding and wiring}

After standard cleaning (acetone cleaning, IPA rinse, DI water
rinse, piranha cleaning, and DI water rinse) in clean room, one
Si(111) piece and one Si/SiO$_2$ remote gate piece are transferred
into an oxygen-free glove box with a typical O$_2$ concentration of
2 ppm. The Si(111) piece is put in a diluted solution of HF/H$_2$O
(1:20) for 2 min to remove the sacrificial SiO$_2$ layer, and then
immersed in a 40$\%$ NH$_4$F solution for 10 min to create an
atomically flat H-passivated Si(111) surface, with DI water rinse
after each step. It is crucial that each of these solutions is
oxygen-free to ensure the high quality of the final device
\cite{Wade97}. Usually they are deoxygenated with a constant
agitation from a magnetic stir rod for 72 hours in the glove box.
Finally, these two pieces are transferred into a vacuum chamber, and
pushed against a sapphire boss to apply a pressure between them,
sufficient to initiate bonding through van der Waals forces .

The bonded device is loosely placed in a customized holder machined
from polyimide material, which is designed to facilitate handling of
the device without applying stress. It is wired in a nitrogen glove
box by hand soldering using indium at 320 $^\circ$C and installed on
a standard 8-pin dual in-line package (DIP) header, which is
compatible with sample holders at the National High Magnetic Field
Laboratory (NHMFL). If only the 2DES is to be investigated, eight
contacts are sufficient, including two gate contacts and six Ohmic
contacts to the 2DES. There are 12 wires in our cryostat. In order
to measure both the 2DES and the 2DHS, the 8 pin DIP header is
inserted into another 20 pin DIP header, and four Ohmic contacts to
the 2DHS are directly wired to the 20 pin DIP header. Two p$^+$
Ohmic contacts 3, 9 are not wired. The wired sample is transferred
from the nitrogen glove box to the cryostat, and the cryostat is
pumped to vacuum ($p<10^{-3}$ mbar) in less than one hour. Figure
\ref{fig:realdevice}(d) shows one of the final devices without
wiring. Similar fabrication processes have been discussed elsewhere
in detail \cite{KottPhd12, McfarlandPhD10}.

\section{\label{sec:characterization}DEVICE CHARACTERIZATION}

\subsection{Topography}

\begin{figure}
 \includegraphics{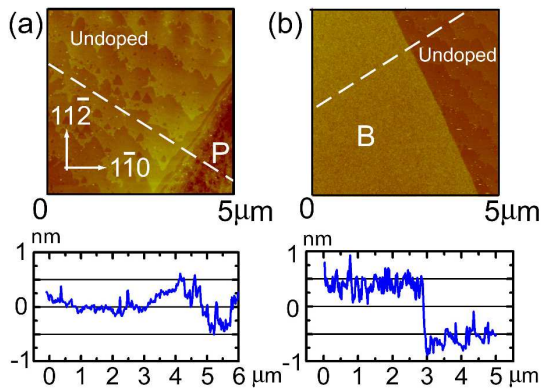}%
 \caption{\label{fig:AFM}AFM images of a H-Si(111) surface with
 (a) a phosphorus-doped contact edge and (b) a boron-doped contact edge.
 They also show atomic steps in undoped areas and the crystallographic
 directions. Bottom graphs show height profiles along white dash lines
 in the top graphs.}
 \end{figure}

Because the Si(111) piece and the Si/SiO$_2$ remote gate piece are
contact bonded through van der Waals forces, successful bonding
requires that the surfaces are clean, flat and smooth. Atomic force
microscopy (AFM) is used to investigate the topography of the
Si(111) surface, particularly the differences in height and flatness
of doped and undoped areas. Figure \ref{fig:AFM} shows that the
Si(111) surface is clean with atomic steps. Typically the doped
contacts on the Si(111) surface are at different elevations from the
undoped Si(111) surface after wet chemical cleaning and passivation.
Specifically, n$^+$ Ohmic contacts (phosphorus doped) are $\sim$ 1
nm below the undoped region, as shown in Fig.\ \ref{fig:AFM}(a);
p$^+$ Ohmic contacts (boron doped) are $\sim$ 1.5 nm above the
undoped region, as shown in Fig.\ \ref{fig:AFM}(b). We have found
that the RCA-1 clean with a H$_2$O$_2$/NH$_4$OH/H$_2$O solution
etches the SiO$_2$ region exposed to ion implantation much faster
than the unimplanted oxide region. Once it etches through the
sacrificial oxide layer completely, it etches the underlying doped
silicon. This results in larger height difference, which may impair
the Ohmic contact to the 2DS. Therefore, the time in the RCA-1 clean
should be minimized, and under 20 minutes \cite{McfarlandPhD10}.
Piranha clean with a H$_2$SO$_4$/H$_2$O$_2$ solution does not have
this issue and thus is preferred when cleaning Si(111) pieces.

\subsection{Gate leakage}

\begin{figure}
 \includegraphics{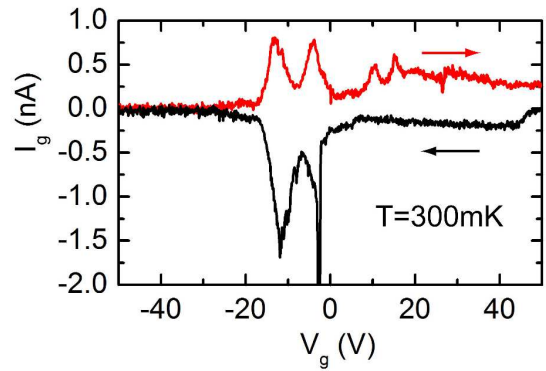}%
 \caption{\label{fig:Gateleak}Gate current $I_g$ vs. gate voltage
 $V_g$ on sample 198 when $V_g$ is swept between -50 V
 ($p_{2d}=7.6\times10^{11}$ cm$^{-2}$) and 50 V ($n_{2d}=7.8\times10^{11}$ cm$^{-2}$)
 at a sweep rate of 0.2 V/s with all wired n$^+$, p$^+$ contacts grounded.
 Arrows indicate the voltage sweep direction. Current peaks at
 $V_g$ $\sim$ -12 V and -3 V are due to
 the formation or depletion of 2DSs. $I_g \sim 0.5$ nA is from the gate
 capacitance.}
 \end{figure}

Samples are intended for measurements at low temperature;
consequently even sub-nanoampere gate leakage can potentially affect
the data. All electrical measurements made in this section
(Sec.~\ref{sec:characterization}) were performed at $T=300$ mK in a
helium 3 refrigerator (Oxford Instruments, HelioxVL). Gate leakage
was checked when all Ohmic contacts were grounded and the gate
voltage V$_g$ was swept between -50 V ($p_{2d}=7.6\times10^{11}$
cm$^{-2}$) and 50V ($n_{2d}=7.8\times10^{11}$ cm$^{-2}$) at a sweep
rate of 0.2 V/sec. The measured gate current was less than 0.5 nA at
-50 V $< V_g <$ -20 V and 0 V $< V_g <$ 50 V, as shown in Fig.\
\ref{fig:Gateleak}. (The capacitance between the gate and the ground
was measured to be about 3 nF, so the gate current of 0.6 nA was
expected.) The static leakage current was less than 50 pA at
$V_g=\pm 50$ V. There is no measurable leakage current for these
devices. In Fig.\ \ref{fig:Gateleak}, the peak current at $V_g$
$\sim$ -12 V and -3 V is due to the formation or depletion of the
2DES or 2DHS.

\subsection{PN junction isolation and leakage}

\begin{figure}
 \includegraphics{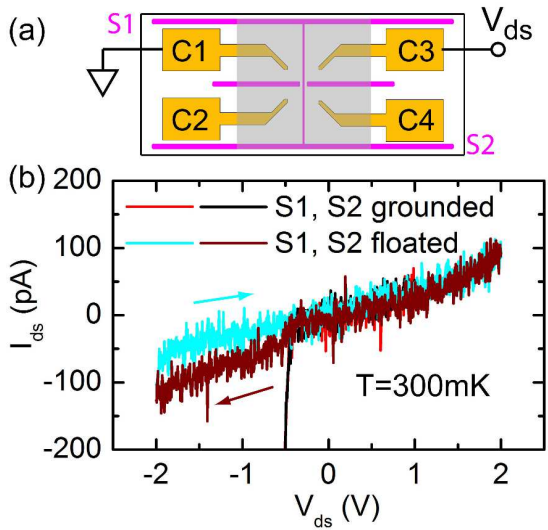}%
 \caption{\label{fig:PNisolation}(a) A test device with four n$^+$ Ohmic
 contacts (yellow) and p$^+$ isolation lines (purple). A center vertical
 p$^+$ isolation line separates the 2DES (shaded area) into two halves.
 (b) Leakage current $I_{ds}$ vs. $V_{ds}$ at $n_{2d}=6\times10^{11}$ cm$^{-2}$
 with either p$^+$ isolation lines (S1, S2) grounded or floated. The leakage due
 to forward biasing happens at $V_{ds}<-0.4$ V when S1, S2 are grounded.
 Arrows show the voltage sweep direction.}
 \end{figure}

In earlier generation devices, a silicon-on-insulator (SOI) remote
gate piece was used to define the 2DS on the H-Si(111) surface with
a shield layer to restrict the electric field of the gate
\cite{Eng05APL}. With the introduction of the Si/SiO$_2$ global
gate, the electric field is all over the H-Si(111) surface, as shown
in Fig.1 (c) (d). PN junction isolation and trench isolation are
used to confine the 2DS. Using the SUPREM-IV process simulator, the
depth of the p$^+$ and n$^+$ Ohmic contacts, defined as the region
where doping concentration is higher than that of the 3D MIT ($\sim
4\times 10^{18}$ cm$^{-3}$ \cite{Rosenbaum80, Dai91}), is determined
to be about 100 nm, as shown in Fig.\ \ref{fig:Dopingprofile}(a)(b).
For the trench isolation, 200 nm deep and 10 $\mu$m wide trenches
are used. Since the depth of the 2DS is less than 10 nm for 2D
carrier density higher than 10$^{11}$ cm$^{-2}$ in the H-Si(111)
vacuum FET \cite{Hwang07, Hwang13}, both PN junction isolation and
trench isolation should work well.

\begin{figure}
 \includegraphics{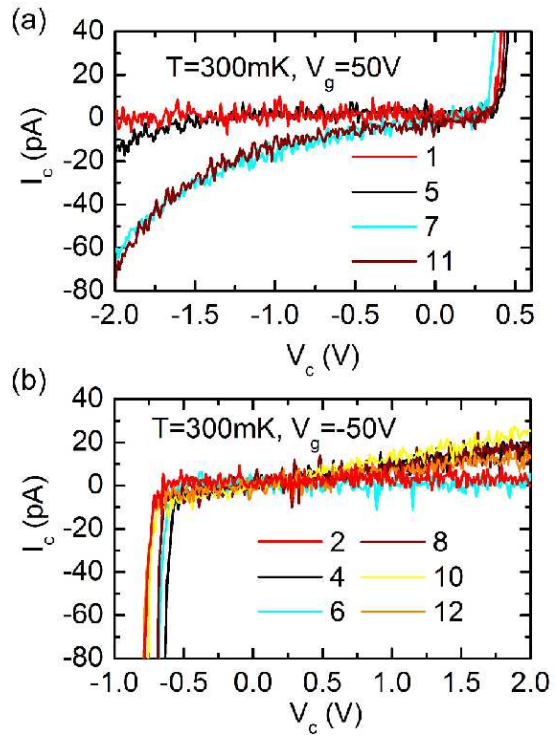}%
 \caption{\label{fig:PNleak} PN junction leakage current on sample 198.
 (a) P$^+$ contact leakage current $I_c$ at
 $V_g=50$ V ($n_{2d}=7.8\times10^{11}$ cm$^{-2}$) when a voltage $V_c$
 is applied at each wired p$^+$ contact, while all n$^+$ contacts are grounded.
 (b) N$^+$ contact leakage current $I_c$ at $V_g=-50$ V
 ($p_{2d}=7.6\times10^{11}$ cm$^{-2}$) when $V_c$ is applied at each
 $n^+$ contact, while all wired p$^+$ contacts are grounded. In both cases,
 the resistance is larger than 100 G$\Omega$ at $-0.5<V_{c}<0.4$ V.}
 \end{figure}

A test device [Fig.\ \ref{fig:PNisolation}(a)] was fabricated to
verify the effectiveness of the PN junction isolation. A 2DES
[shaded area in Fig.\ \ref{fig:PNisolation}(a)] is separated by a 10
$\mu$m wide boron ion-implanted line (the vertical line). At an
electron density of $6\times10^{11}$ cm$^{-2}$, the resistance
between n$^+$ contacts C1 and C3 were measured, while p$^+$ contacts
S1 and S2 were either grounded or floated. In both cases, Fig.\
\ref{fig:PNisolation}(b) shows that the resistance between C1 and C3
was larger than 25 G$\Omega$ at $-0.2<V_{ds}<1$ V, likely dominated
by inter-wire leakage, so the PN junction isolation is effective.
When S1 and S2 were grounded, the PN junction began to leak at
$V_{ds}=-0.4$ V because of forward bias.

For the device under investigation (sample 198), we also measured
the PN junction leakage current. At an electron density of
$7.8\times 10^{11}$ cm$^{-2}$, the leakage current from p$^+$
contacts 1, 5, 7, 11 to the 2DES was measured by applying a voltage
$V_c$ at each contact, while all n$^+$ contacts were grounded. (Here
 p$^+$ contacts 3, 9 were not wired, and thus floated.) The resistance
 was larger than 100 G$\Omega$ at
$-0.5<V_{c}<0.4$ V, as shown in Fig.\ \ref{fig:PNleak}(a).
Similarly, the leakage current from n$^+$ contacts 2, 4, 6, 8, 10,
12 to the 2DHS was measured by applying a voltage $V_c$ at each
n$^+$ contact, while all wired p$^+$ contacts were grounded. Figure
\ref{fig:PNleak}(b) shows that the resistance was also larger than
100 G$\Omega$ at $-0.5<V_{c}<1$ V. In our transport measurements,
lateral confining gates were floated with respect to the 2DS, and
the source-drain voltage is less than $\pm 0.1$ V with the
source-drain current equal or larger than 10 nA, so the leakage
current from the PN junction isolation has only a negligible effect
on the measurements.

We also measured the conductance between n$^+$ contacts C1 and C2
while negatively biasing p$^+$ contacts S1 and S2. The results show
that the conductance decreases with increasing negative bias
voltage, which suggests that depletion is occurring and reverse
biasing lateral gates to confine carriers should be possible.

\subsection{Ohmic contacts}

Contacts to the 2DSs on H-Si(111) surfaces have been problematic
especially at low temperature ($<1$ K) and low carrier density,
which limits the lowest accessible 2D carrier density in
experiments. The main issues are large contact resistances (on the
order of 1 M$\Omega$) and nonlinearity in current-voltage (IV)
curves \cite{KottPhd12}.

For the device without STI, the contact problem can be mitigated by
the presence of SiO$_2$ over the contact regions. At the center 2DS
region, the dielectric is vacuum with a dielectric constant of 1; at
the contact regions, SiO$_2$ is the dielectric with a dielectric
constant of 3.9. So at the same gate voltage, the contact regions
have about four times higher carrier density than the center 2DS
region when the threshold voltages are negligible, and the contact
resistance is reduced.

\begin{figure}
 \includegraphics{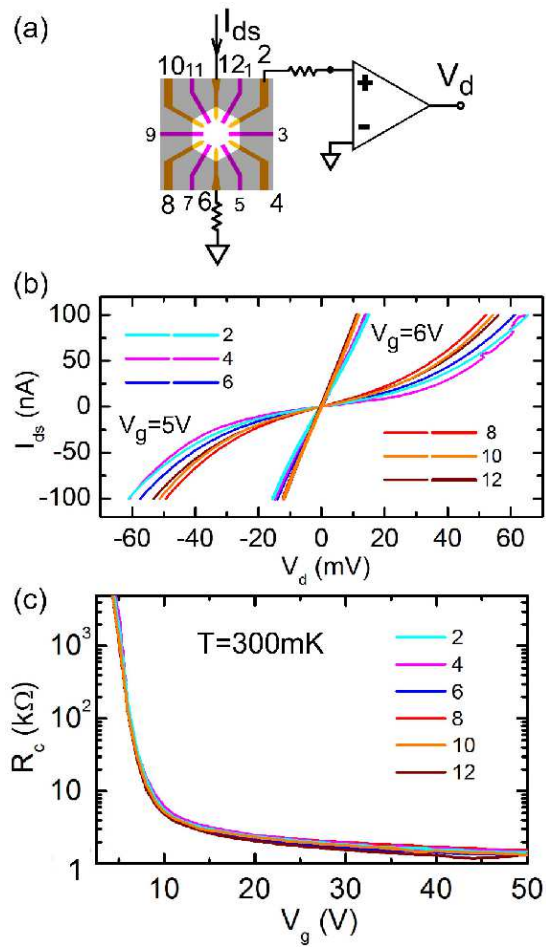}%
 \caption{\label{fig:Ncontact} N$^+$ contacts on sample 198 measured at 300 mK. (a) Circuit diagram measuring contact 6,
 with current $I_{ds}$ injected from contact 12 to contact 6 and voltage $V_d$ measured
 between contacts 2 and 6. (b) Current-voltage curves of n$^+$ contacts at
 $V_g=5$ V ($n_{2d}=0.87\times10^{11}$ cm$^{-2}$)) and 6V ($n_{2d}=1.02\times10^{11}$ cm$^{-2}$).
 (c) Contact resistance $R_c$ as a function of gate
 voltage $V_g$. At $V_g<6$ V, $R_c$ is calculated when $|I_{ds}|\leqslant4$ nA; otherwise, $R_c$ is
 determined at $|I_{ds}|\leqslant100$ nA.}
 \end{figure}

The contact resistance was measured at $T=300$ mK, using a homemade
voltage amplifier with input bias current $\sim$ 1 pA and input
impedance $> 200$ G$\Omega$. Figure \ref{fig:Ncontact}(a) shows the
circuit diagram measuring contact 6, where current $I_{ds}$ is
injected from contact 12 to contact 6, and voltage $V_d$ is measured
between contacts 2 and 6. The IV curves were nonlinear at $V_g=5$ V
($n_{2d}=0.87\times10^{11}$ cm$^{-2}$) and became linear at $V_g=6$
V ($n_{2d}=1.02\times10^{11}$ cm$^{-2}$), as shown in Fig.\
\ref{fig:Ncontact}(b). So the lowest accessible 2D electron density
was about $0.9\times10^{11}$ cm$^{-2}$ at 300 mK for this device.
Figure \ref{fig:Ncontact}(c) shows contact resistance $R_c$ as a
function of gate voltage $V_g$. The sheet resistance of the n$^+$
regions was measured to be 129 $\Omega/\square$ at 4.2 K, and this
3D metallic resistance should not change too much at 300 mK because
of the lack of phonon scattering with the doping concentration well
above the 3D MIT density. The n$^+$ contact resistance is calculated
to be about $2.5$ k$\Omega$ if only considering this sheet
resistance, which is much smaller than the measured contact
resistance at $V_g<10$ V. The commercial software package COMSOL
Multiphysics \cite{[{http://www.comsol.com/}][{}]comsol} is used to
calculate the contact resistance including both the n$^+$ contact
regions and the 2DS region. For example, at $V_g=10$ V, the sheet
resistances of the n$^+$ contact regions and the 2DES region are 129
$\Omega/\square$ and 807 $\Omega/\square$ respectively, and the
calculated contact resistance is about 2.9 k$\Omega$, which is
significantly less than measured contact resistance of 4.9
k$\Omega$. The difference is likely related to the highly resistive
transition region between the metallic n$^+$ contact and the 2DES,
where the doping concentration is less than the 3D MIT density
($\sim 3.74\times 10^{18}$ cm$^{-3}$ \cite{Rosenbaum80}), as shown
in Fig.\ \ref{fig:Dopingprofile}(a). The 2DES in the transition
region can undergo a 2D MIT and become an insulator at low carrier
density and low temperature, resulting in high contact resistance
\cite{Kravchenko04}.

\begin{figure}
 \includegraphics{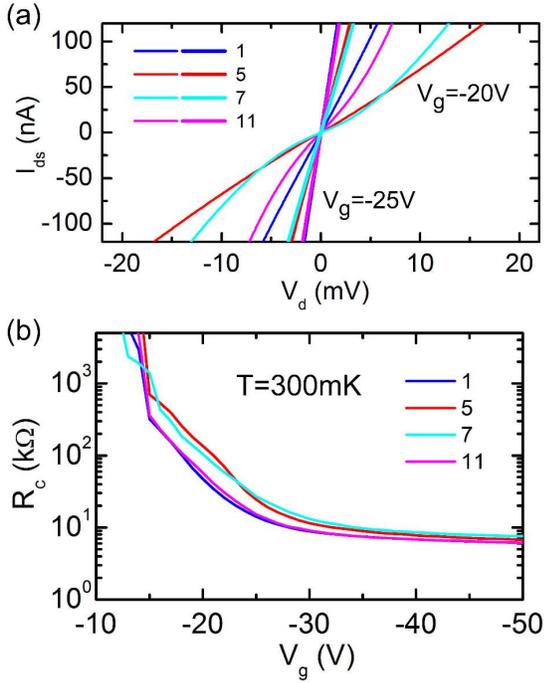}%
 \caption{\label{fig:Pcontact} P$^+$ contacts on sample 198 measured at 300 mK.
 (a) Current-voltage curves of p$^+$ contacts
 at $V_g=-20$ V ($p_{2d}=2.94\times10^{11}$ cm$^{-2}$) and -25 V
 ($p_{2d}=3.72\times10^{11}$ cm$^{-2}$).
 (b) Contact resistance $R_c$ as a function of gate voltage
 $V_g$. At $-V_g<25$ V, $R_c$ is calculated
 when $|I_{ds}|\leqslant4$ nA; otherwise, $R_c$ is
 determined at $|I_{ds}|\leqslant100$ nA.}
 \end{figure}

The p$^+$ contact resistance was also measured. Figure
\ref{fig:Pcontact}(a) shows that the IV curves were nonlinear at
$V_g=-20$ V ($p_{2d}=2.94\times10^{11}$ cm$^{-2}$) and became linear
at $V_g=-25$ V ($p_{2d}=3.72\times10^{11}$ cm$^{-2}$), which is much
higher than the corresponding gate voltage for the n$^+$ contacts.
The lowest accessible 2D hole density was about $3\times10^{11}$
cm$^{-2}$ at 300 mK. Figure \ref{fig:Pcontact}(b) shows contact
resistance $R_c$ as a function of gate voltage $V_g$. We measured
the sheet resistance of the p$^+$ regions to be 103 $\Omega/\square$
at 4.2 K, which is comparable to the n$^+$ regions. If we compare
the doping profile of the phosphorus doped region and the boron
doped region (Fig.\ \ref{fig:Dopingprofile}), especially the top
$\sim10$ nm area where 2DSs are populated, there is no significant
difference. It is quite puzzling why the p$^+$ contacts are much
worse than the n$^+$ contacts.

\begin{figure}
 \includegraphics{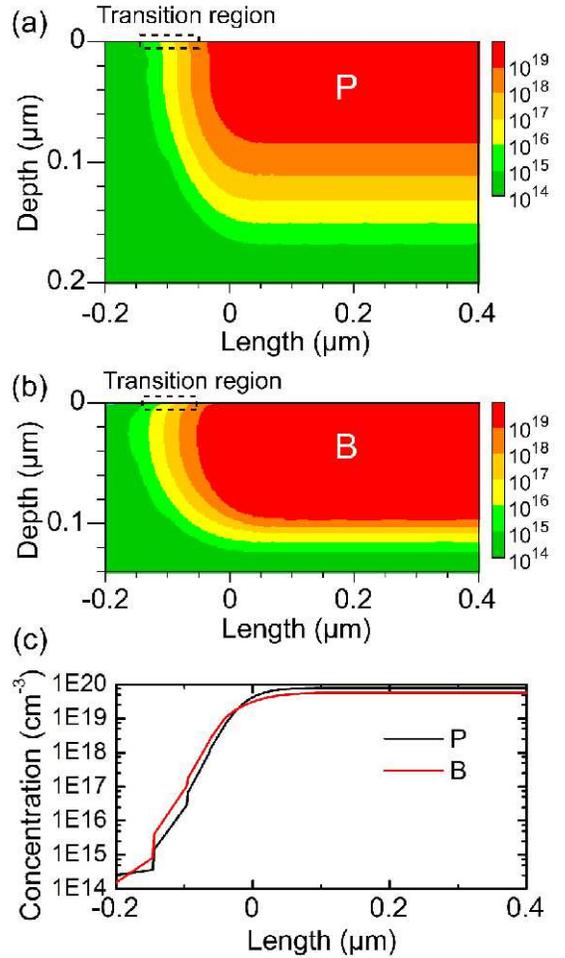}%
 \caption{\label{fig:Dopingprofile}2D doping profile of (a) phosphorus-doped
 region and (b) boron-doped region in Si using SUPREM-IV simulator.
 In (a), $4.5\times10^{14}$cm$^{-2}$, 50 keV phosphorus ions are implanted through a 30 nm oxide;
 in (b), $9\times10^{14}$cm$^{-2}$, 15 keV boron ions are implanted through a 30 nm oxide.
 Both are annealed at 950 $^\circ$C for 1 min. Ions are implanted at 0$^\circ$ with
 vertical photoresist edge at Length=0. (c) Doping concentrations of phosphorus and boron
 along the Si surface at (a) and (b).}
 \end{figure}

There are probably two reasons. One is that electrons have much
stronger Coulomb disorder screening strength than holes due to the
very different valley degeneracy (6:1) \cite{Hu15PRL}; the other is
that there could be very different lateral defect profiles, such as
end-of-range (EOR) dislocation loops \cite{Plummer00}. The very
different defect profiles can be caused by two reasons: first,
phosphorus and boron ions have very different critical doses for
amorphization of silicon \cite{Morehead70}. At room temperature, the
critical dose of phosphorus ions is $\sim  10^{15}$ cm$^{-2}$, about
twice as high as the dose used here; while it is $\sim 10^{17}$
cm$^{-2}$ for boron ions, which is about 100 times higher than the
current dose. It is often easier to regrow the crystal from an
amorphous layer via solid state epitaxy (activation energy $\sim
2.3$ eV in Silicon) than it is to anneal out defects (activation
energy $\sim 5$ eV). Thus, two schemes for ion implantation are
usually used: either perform ion implantation above the critical
dose and use low temperature annealing to regrow material or perform
ion implantation below the critical dose and use high temperature
annealing to get rid of defects \cite{Plummer00}. Second, phosphorus
and boron ions have different mask edge effects, such as lateral
straggle and shadowing effect, resulting in different lateral doping
profile \cite{Furukawa72, Privitera95}.

It is probably beneficial to do the ion implantation at higher dose
and cryogenic temperature. We observed improvement of the p$^+$
contacts when the dose of the boron ion implantation was increased
from $6\times10^{14}$cm$^{-2}$ to $2.4\times10^{15}$cm$^{-2}$, then
the contacts were annealed at 1000 $^\circ$C for 10 min
\cite{Hu10APL}. But there is a possibility that the contact bonding
could be difficult, because of the larger height difference between
the p$^+$ contact region and the undoped region after wet chemical
etching than that shown in Fig.\ \ref{fig:AFM} (b). In order to
access lower 2D carrier densities, a structure like a bi-layer gate
structure using SOI, as we will discuss in
Sec.~\ref{sec:degradation}, can be used to solve the contact
problem.

\section{\label{sec:transport}TRANSPORT MEASUREMENTS}

\subsection{2D carrier densities vs. gate voltage}
\begin{figure}
 \includegraphics{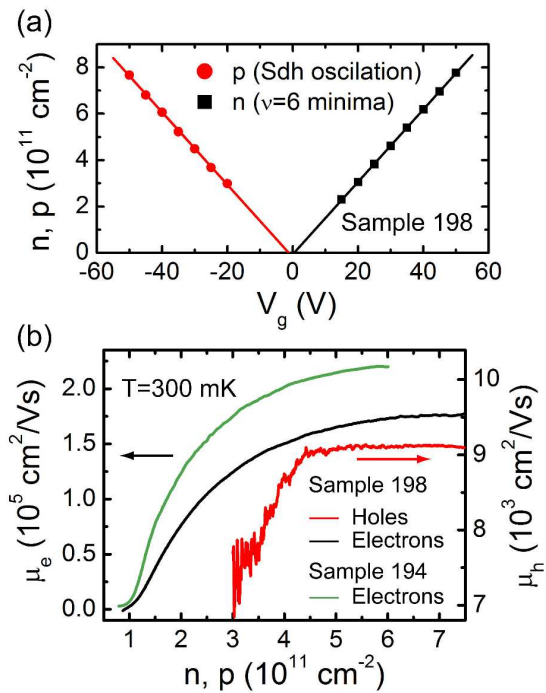}%
 \caption{\label{fig:DensitymobVg}(a) Electron and hole densities as
 a function of gate voltage for sample 198. The electron density is
 determined from the magnetic field of $R_{xx}$ minima at the filling factor
 $\nu=6$, and the hole density is calculated from the Shubnikov-de Haas
 oscillations. (b) 2D carrier mobility of sample 194 and 198 vs. carrier
 density at 300 mK. For the ambipolar device 198, the ratio between the
 peak electron and hole mobilities is about 20.}
 \end{figure}

Longitudinal ($R_{xx}$) and Hall ($R_{xy}$) resistances were
measured at 300 mK using standard low-frequency AC lock-in
techniques with an excitation current of 100 nA at 7 Hz. Sheet
resistance was determined by standard Van der Pauw method.
Magnetotransport measurements were performed at different gate
voltages in a perpendicular magnetic field ($B_\perp$) up to 12 T.
From the SdH oscillations, the 2D hole densities are determined at
each gate voltage. The 2D electron densities $n_{2d}$ are calculated
from the magnetic field $B$ of $R_{xx}$ minima at the filling factor
$\nu=6$: $n_{2d}=\nu eB/h$, where $e$ is the electron charge, and
$h$ is Planck's constant. The 2D densities and their linear fits for
sample 198 are shown in Fig.\ \ref{fig:DensitymobVg}(a). For
electrons, the slope is $1.563\pm0.004\times10^{10}$ cm$^{-2}$/V,
and the threshold voltage is $V_{e\_th}=0.4\pm0.1$ V. For holes, the
slope is $1.566\pm0.001\times10^{10}$ cm$^{-2}$/V, and the threshold
voltage is $V_{h\_th}=-1.21\pm0.03$ V. If a parallel plate capacitor
model is used, the equivalent depth of the vacuum cavity is
$355\pm1$ nm, which is consistent with the measurement result using
a profilometer. The interface trap density in the band gap is
determined to be:
$D_{it}\approx(V_{e\_th}-V_{h\_th}-E_{gap})\times1.56\times10^{10}\approx0.8\times10^{10}$(cm$^{-2}$),
where the band gap $E_{gap}$ is 1.17 V for silicon at 0 K
\cite{sze81}.

\subsection{2D carrier mobilities}

2D carrier mobilities are calculated from the sheet resistance, and
plotted in Fig.\ \ref{fig:DensitymobVg}(b). For sample 194, the peak
electron mobility was $2.2\times10^5$ cm$^{2}$/Vs at 300 mK. For
sample 198, the peak electron mobility and the peak hole mobility
were very different, $1.76\times10^5$ cm$^{2}$/Vs and
$9.1\times10^3$ cm$^{2}$/Vs respectively at 300 mK, although the
electrons and holes were populated in the same conduction channel.
The ratio between the peak electron and hole mobilities is about 20,
which cannot be explained by the different effective mass of
electrons and holes. In fact, the effective mass is comparable
($\sim0.38 m_e$ for electrons \cite{Neugebauer75, Abstreiter76,
Shashkin07} and $\sim0.34 m_e$ for holes \cite{Kotthaus77}, where
$m_e$ is the electron mass) at the investigated densities. This is
quite different from the results of GaAs ambipolar devices, in which
the ratio of the peak electron mobility to the peak hole mobility is
about 5, consistent with the different effective mass of electrons
and holes \cite{Chen12APL}. The large difference in the electron and
hole mobilities is a direct consequence of the very different
Coulomb disorder screening strength due to the different valley
degeneracies of electrons ($g_v=6$) and holes ($g_v=1$) at the
Si(111) surface \cite{Hu15PRL}.

In addition, the electron mobility increases monotonically with the
electron density, and the mobility saturation is not observed up to
the highest electron density measured ($n=7.8\times10^{11}$
cm$^{-2}$), which is limited by the gate breakdown voltage. From the
electron conductivity $\sigma$ vs. electron density $n$ data, the
fitting exponent $\alpha$ in the relation of $\sigma \sim n^\alpha$
is about 1.3 at $n=7\times 10^{11}$ cm$^{-2}$  in these devices
\cite{Hu15PRL}, which is consistent with the case ($\alpha \approx
1.3$) where the electron mobility is mainly limited by 3D bulk
impurities, while deviating from the case ($\alpha \approx 1.05$)
where 2D interface impurities limit the electron mobility
\cite{[{Internal communication with E. H. Hwang and S. Das Sarma;
see also}][{}] Sarma13}. We have developed a rebonding technique,
which can revive a sample by taking it apart, cleaning the two
pieces and H-passivating the Si(111) piece, then rebonding them
together. The piranha cleaning and the diluted HF etching can remove
the top few nm of silicon from the Si(111) piece. We tested it on
sample 195 with an initial electron mobility of $5.6\times10^4$
cm$^{2}$/Vs at $n=6\times10^{11}$ cm$^{-2}$ and $T=300$ mK. After
repeated piranha cleaning and diluted HF etching three times, the
electron mobility increased to $7.2\times10^4$ cm$^{2}$/Vs
($n=6\times10^{11}$ cm$^{-2}$, $T=300$ mK), which is likely due to
the removal of the surface layers with higher 3D impurity density
from earlier processes. All these suggest that the electron mobility
is still mainly limited by 3D bulk impurities for current generation
devices.

\subsection{Magnetotransport measurements}

\begin{figure}
 \includegraphics{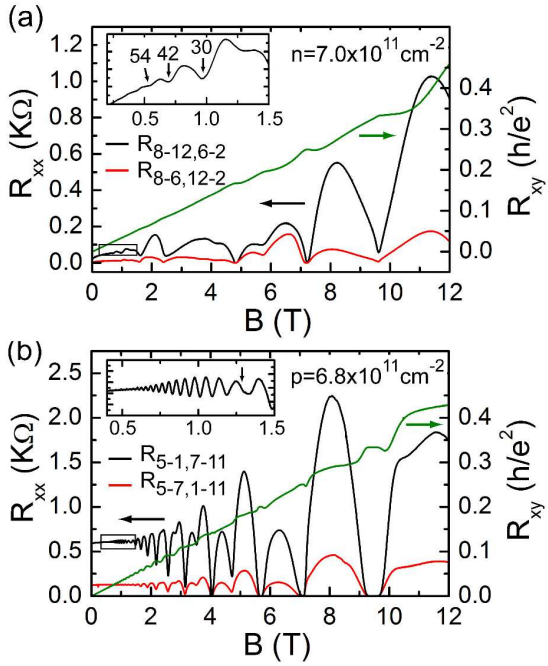}%
 \caption{\label{fig:Magnetotransport} Magnetoresistance and Hall
 resistance measured at 300 mK for similar (a) electron and (b) hole densities
 on sample 198. The frequency of the SdH oscillations is much higher
 for holes than electrons owing to different valley degeneracy (1:6).
 The insert in (a) clearly shows the 12-fold periodicity from the sixfold
valley degeneracy of the 2DES and the twofold spin degeneracy. The
insert in (b) shows a beating pattern with a beating node at the
arrow. }
 \end{figure}

All magnetotransport measurements were carried out in $B_\perp$ in
this paper. Figure \ref{fig:Magnetotransport} shows the
magnetoresistance for similar electron and hole densities in the
same device (sample 198) at $T=300$ mK. The frequency of the SdH
oscillations was much higher for holes than electrons at low field,
because electrons and holes have different areas of the Fermi
surface resulting from the different valley degeneracies. For the
2DHS, the characteristic beating node was observed at $B=1.25$ T for
$p_{2d}=6.8\times10^{11}$ cm$^{-2}$ \cite{Hu10APL}, and the IQHE was
also observed at high field ($B>5$ T). For the 2DES, the SdH
oscillations began at $B\approx0.25$ T, and $R_{xx}$ minima showed
clearly the 12-fold periodicity [the insert of Fig.\
\ref{fig:Magnetotransport}(a)] from the sixfold valley degeneracy
and the twofold spin degeneracy. In contrast to previous
investigations of Si(111) transport on MOSFETs with peak electron
mobility $\mu\leqslant2,500$ cm$^{2}$/Vs which have shown
conflicting valley degeneracies of two \cite{Shashkin07} and six
\cite{Tsui79}, all six current generation H-Si(111) devices with
electron mobility of more than 50,000 cm$^{2}$/Vs show exclusively
the sixfold valley degeneracy at low magnetic field.

\subsection{Three magnetoresistance measurements with threefold rotational symmetry
}

\begin{figure}
 \includegraphics{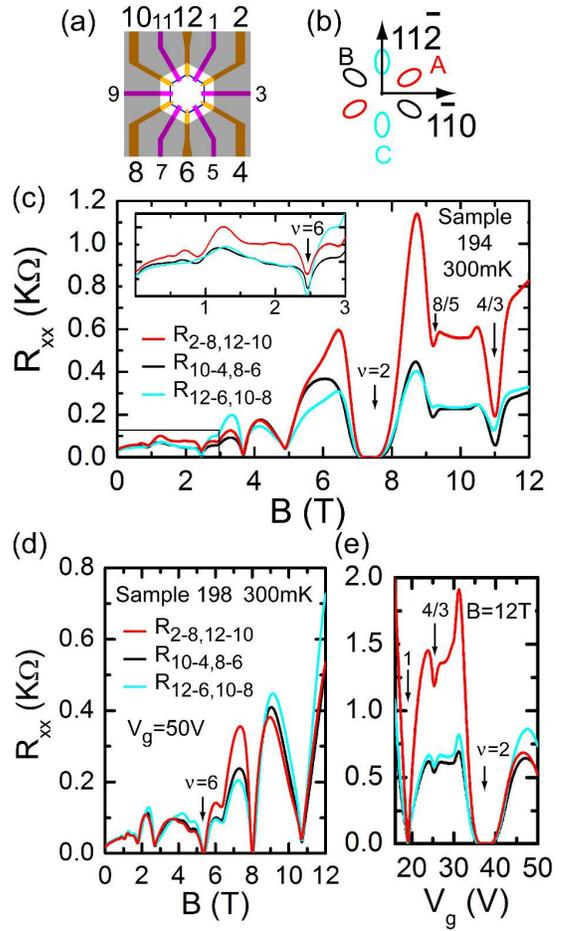}%
 \caption{\label{fig:3MagnetoR} (a) Top view of a device with six
electron contacts and six hole contacts in relation to (b) the 3
pairs of the valleys, labeled as A, B and C. (c) Three
magnetoresistance measurements on sample 194 at
$n_{2d}=3.6\times10^{11}$ cm$^{-2}$ with the insert showing the low
field behavior. (d) (e) Three magnetoresistance measurements on
sample 198 at $n_{2d}=7.8\times10^{11}$ cm$^{-2}$. For both samples
194 and 198, one trace is higher than the other two similar traces
at $1<\nu<2$.}
 \end{figure}

Because electrons have six equivalent valleys on the Si(111)
surface, the six n$^+$ Ohmic contacts are placed along the major
axes of the constant energy ellipses, labeled as 2,4,$\dots$, 12, as
shown in Fig.\ \ref{fig:3MagnetoR} (a)(b). The six equivalent
valleys form three pairs with opposite momentum, labeled as A, B and
C. The degeneracy of valley pairs with $\pm \vec{k}$  symmetry on
the Si(111) surface cannot be broken within the effective mass
approximation or by a confinement potential \cite{Stern67,
Rasolt86}. Similarly, longitudinal magnetoresistance $R_{xx}$,
measured with current contacts on opposite sides of the hexagonal
devices, also has three different directions defined by the
direction of current flow. Here four-terminal resistance is defined
as $R_{i-j,l-m}=V_{l-m}/I_{i-j}$ (i, j, l, m = 2,4,$\dots$, 12),
where current $I_{i-j}$ is injected from contact i to contact j, and
voltage $V_{l-m}$ is measured between contacts l and m. If the
current flows from 2 to 8, it is called trace A ($R_{2-8,12-10}$).
If the current flows from 10 to 4, it is called trace B
($R_{10-4,8-6}$), and if the current flows from 12 to 6, it is
called trace C ($R_{12-6,10-8}$). These three magnetoresistances
$R_{xx}$ with threefold rotational symmetry can be used to determine
the underlying symmetry of the 2DES, and help identify the valley
occupancies of the 2DES. For example, if electrons only occupy
valley pair A, trace A will show higher resistance, while trace B
and C will show similar lower resistance, due to the different
effective mass.

Figure \ref{fig:3MagnetoR}(c)-(e) show the three magnetoresistance
measurements on sample 194 and 198. Data were taken at 300 mK and a
density of $3.6\times10^{11}$ cm$^{-2}$ or $7.8\times10^{11}$
cm$^{-2}$ for samples 194 and 198 respectively. The two simplest
cases are $B=0$ and $B$ at $\nu=3/2$ where composite fermions (CFs)
experience a zero effective magnetic field. At $B=0$, both devices
show more or less isotropic resistances. For sample 194, the three
resistances are 37.2 $\Omega$, 30.4 $\Omega$, and 30.3 $\Omega$; for
sample 198, they are 16.0 $\Omega$, 17.9 $\Omega$, and 17.9
$\Omega$, as shown in Fig.\ \ref{fig:3MagnetoR}(c)(d). At $B_{3/2}$,
the devices are anisotropic. For sample 194, the three
magnetoresistances are 558.1 $\Omega$, 226.3 $\Omega$, and 232.8
$\Omega$. For sample 198, they are 1356.9 $\Omega$, 615.4 $\Omega$,
and 664.7 $\Omega$. The observed anisotropy can be explained by CFs'
occupying valley pair A. Because of the effective mass anisotropy,
trace A shows higher resistance, and traces B and C show similar
lower resistance. The ratios between the higher and lower
resistances are 2.43 and 2.12 for sample 194 and 198 respectively.

In a Drude model with noninteracting electrons, assuming an
isotropic scattering time, the resistance ratio is equal to the mass
ratio $m_l/m_t$, where $m_l=0.67m_e$ and $m_t=0.19m_e$
\cite{Stern67} are the longitudinal effective mass and the
transverse effective mass respectively. For CFs, all physical
quantities are determined by the Coulomb interaction \cite{Jain07}.
In an isotropic 2DS, the Coulomb interaction is $\propto
1/\sqrt{x^2+y^2}$, where $x$ and $y$ are components of the distance
between two electrons. According to Ref.~\onlinecite{Gokmen10}, in a
system with an anisotropic Fermi surface, it can be mapped to a
system with an isotropic Fermi surface with an anisotropic Coulomb
interaction $\propto 1/\sqrt{x^2\gamma^2+y^2/\gamma^2}$, where
$\gamma=(m_l/m_t)^{1/4}$, $x$ and $y$ are along the major and minor
axes of the constant energy ellipse. If the transport mass of CFs
along some direction is determined by the strength of the Coulomb
interaction along this direction, the mass-anisotropy ratio of CFs
is given by $\gamma^2=\sqrt{m_l/m_t}$. We have calculated two cases
using COMSOL assuming the isotropic scattering time: in the first
case, the mass-anisotropy ratio of CFs is assumed to be $m_l/m_t$;
in the second case, it is assumed to be $\sqrt{m_l/m_t}$. Using a
hexagonal geometry similar to Fig.\ \ref{fig:schematics}(c) and
assuming the anisotropic resistance solely from the anisotropic
effective mass, the calculated ratios between the higher and lower
resistances are 3.58 and 2.07 respectively for these two cases. The
measured ratios 2.43 and 2.12 fall between these two cases and are
closer to the second case. Our results are similar to
Ref.~\onlinecite{Gokmen10}, but deviate from a theoretical study
which predicts almost isotropic CF effective mass
\cite{Balagurov00}. Because the scattering time anisotropy is not
well known, the above discussion becomes more complicated.
Nevertheless, in a simplest picture, electrons occupy the lowest
valley pair at filling factor $1<\nu<2$. Figure
\ref{fig:3MagnetoR}(c) and (e) show that the corresponding CFs
occupy valley pair A, and qualitatively preserve the valley
anisotropy, in addition to the twofold valley degeneracy manifested
by observed exclusive even numerator fractional quantum Hall (FQH)
states (8/5 and 4/3) \cite{Kott14PRB}.

Moreover, at filling factor $\nu$ between 4 and 6, the devices also
show anisotropic magnetoresistances. For sample 194, trace C is
highest, while trace A is highest for sample 198. Both cases can be
qualitatively explained when the Fermi level is located at valley
pair C or valley pair A respectively. For filling factor $\nu$
between 3 and 4, the three magnetoresistances are similar for both
devices, indicating an isotropic distribution of electrons in the
three valley pairs. This cannot be explained by a valley occupancy
of noninteracting electrons, and may related to the quantum Hall
nematic phase \cite{Abanin10, Kumar14}. When filling factor $\nu$ is
between 2 and 3, the three magnetoresistances are quite different
for sample 194, but similar for sample 198. At lower B ($\nu>6$),
the three magnetoresistances are similar especially for sample 198,
consistent with the approximation that six valleys are equally
occupied.

The three magnetoresistance measurements indeed provide a powerful
tool to investigate the underlying symmetry of the 2DES on the
Si(111) surface. In addition, a tilted magnetic field can
re-distribute electrons between the three pairs of valleys through
the application of an in-plane magnetic field \cite{Stern67}. If
these two techniques are combined, it will further our understanding
of this 2DES and the nature of the observed anisotropy.

\subsection{Magnetotransport measurements up to 35 T}

\begin{figure}
 \includegraphics{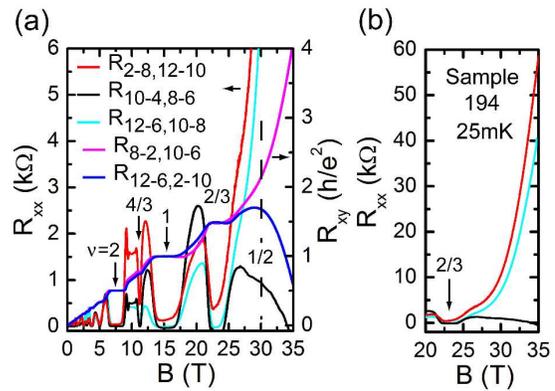}%
 \caption{\label{fig:3MagnetoR35T}(a) Three magnetoresistances
and Hall resistance measured on sample 194 at 25 mK with the clearly
observed 2/3 fractional quantum Hall state. (b) The three
magnetoresistances are highly anisotropic at $\nu<2/3$, with two
similar traces higher than the other one. This may indicate the
change of the underlying symmetry.}
 \end{figure}

Sample 194 was further investigated at filling factor $\nu<1$ in the
portable dilution refrigerator (PDF) at NHMFL with a magnetic field
up to 35 T. The device was measured at $T=25$ mK and density of
$3.6\times10^{11}$ cm$^{-2}$. As shown in Fig.\
\ref{fig:3MagnetoR35T}(a), the 2/3 FQH state is clearly observed
with a well developed hall plateau. Moreover, Fig.\
\ref{fig:3MagnetoR35T}(b) shows that the three magnetoresistances
are highly anisotropic at $\nu<2/3$. It is less than 1 k$\Omega$
along trace B, and about 50 k$\Omega$ along traces A and C. This
50:1 anisotropy cannot arise solely from the different effective
mass. The three magnetoresistances also change from one higher
resistance trace with two similar lower resistance traces at
$\nu>2/3$, to two similar higher resistance traces with one lower
resistance trace at $\nu<2/3$. This may indicate the change of the
underlying symmetry of the 2DES.

\begin{figure}
 \includegraphics{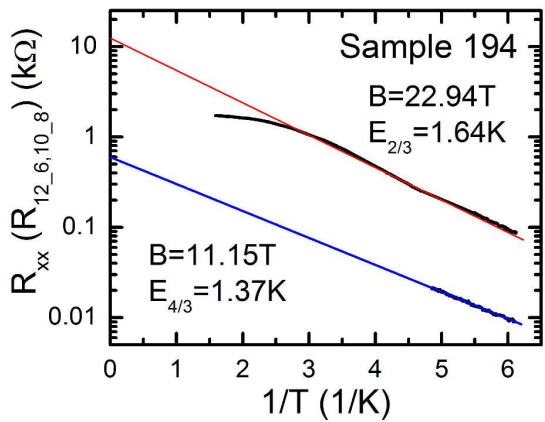}%
 \caption{\label{fig:activationenergy}Arrhenius plots for $\nu=2/3$, 4/3
showing activated behavior on sample 194.}
 \end{figure}

At filling factor $\nu$ between 1 and 2, only even numerator FQH
states (8/5 and 4/3) are observed with no 5/3 FQH state, consistent
with twofold valley-degenerate CFs \cite{Kott14PRB}. It is quite
interesting to find out the underling symmetry of the 2/3 FQH state,
i.e. whether it also has the twofold valley degeneracy. Because
valley degenerate 2/3 FQH state may be related to non-Abelian states
which are being intensively studied for possible applications in
intrinsically fault tolerant quantum computation \cite{Geraedts15,
Peterson15, Zhu15, liu15}, the question becomes more important. We
have determined the activation energies at $\nu=2/3$ and 4/3 from
the temperature dependence of magnetoresistance $R_{xx}$ (Fig.\
\ref{fig:activationenergy}). The activation energies are comparable,
1.64 K for $\nu=2/3$ and 1.34 K for $\nu=4/3$. In a twofold
valley-degenerate system, 4/3 ($=2-2/3$) state can be related to 2/3
state by using particle-hole symmetry \cite{Boeb85, Park01, Jain07,
Kott14PRB}; their activation energies are proportional to
$\sqrt{B}$, and $E_{2/3}/E_{4/3}=\sqrt{22.94/11.15}=1.4$. The
measured ratio is about 1.2, which may imply that the 2/3 FQH state
also has the twofold valley degeneracy.

\section{\label{sec:degradation}DEVICE DEGRADATION}

The H-Si(111) surface is encapsulated in a vacuum cavity in the
H-Si(111) vacuum FET [Fig.\ \ref{fig:schematics}(a)], but the seal
is not perfectly hermetic. Consequently H-Si(111) vacuum FETs still
suffer from device degradation in air owing to degradation of
H-terminated surfaces \cite{Franc97, Royea00, Lauren03}, which may
limit their applications. However, if care is taken to limit the
time of the device spent in air, while keeping it mostly in vacuum
or a nitrogen gas environment, the device degradation is manageable.

\begin{figure}
 \includegraphics{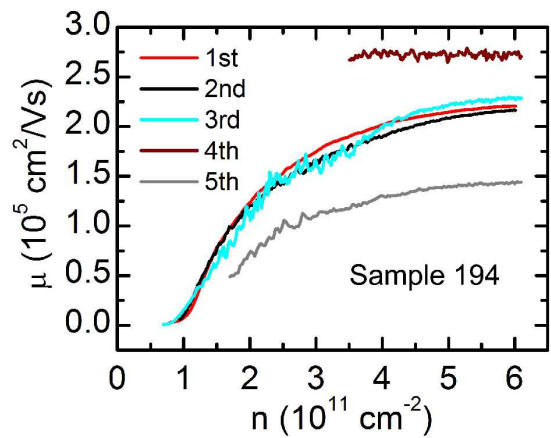}%
 \caption{\label{fig:194mobdegrade}Electron mobility of sample 194
vs. electron density during five cooldowns, measured at 300 mK
except the 4th cooldown (at 25 mK). There was one week of air
exposure between the 4th and 5th cooldowns.}
 \end{figure}

\begin{figure}
 \includegraphics{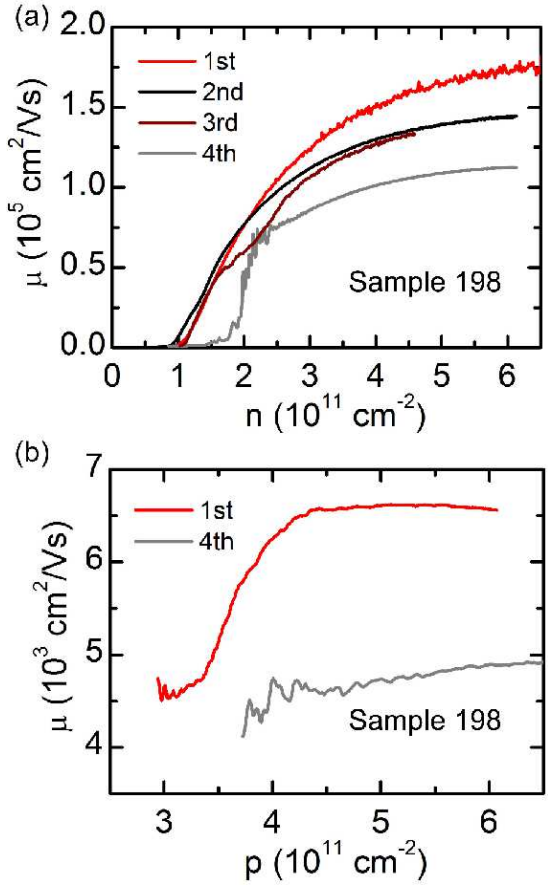}%
 \caption{\label{fig:198mobdegrade}(a) Electron mobility of sample 198
as a function of electron density during four cooldowns, measured at
300 mK except the 3rd cooldown (at 25 mK). One week air exposure
happened between the 3rd and 4th cooldowns. (b) Hole mobility of
sample 198 vs. hole density at 1st and 4th cooldowns, measured at
4.2 K.}
 \end{figure}

The degradation of the two devices has been recorded. Sample 194 was
cooled down five times in a period of 246 days, as shown in Fig.\
\ref{fig:194mobdegrade}. The mobility was mostly measured at 300 mK,
except the fourth cooldown (at 25 mK). As described in
Sec.~\ref{sec:fab}, after the device was bonded and wired in the
nitrogen glove box, it was rapidly loaded into the cryostat at the
first cooldown. There was about one hour exposure in air before
pumping down the cryostat at each cooldown. After the device was
measured for about two weeks, it was warmed up, put into a nitrogen
gas filled barrier foil ziplock storage bag with oxygen absorbers,
and then the bag was heat sealed. After one week, it was cooled down
and measured a second time at day 26. Figure \ref{fig:194mobdegrade}
shows that the change of the peak electron mobility was about 2$\%$.
The device was then warmed up and stored in a nitrogen glove box
(O$_2$ $<10$ ppm) for about 6 months. Right before transferring the
device to the NHMFL, the device was measured a third time. The peak
electron mobility increased by about 3$\%$, which may be due to the
different temperatures or the different contact resistances. So
within 211 days, the device mobility did not change much when
keeping it mostly in a nitrogen gas environment. However, the
mobility fluctuations were pronounced at lower electron densities at
the third cooldown, which was caused by the degradation of the
contacts. For transport to NHFML, the device was again placed in a
nitrogen gas filled storage bag, as described above. After the NHMFL
experiment, the device was exposed in air for about one week, and
the peak electron mobility decreased by 35$\%$ at the fifth
cooldown.

Similarly, sample 198 was cooled down four times in 193 days, as
shown in Fig.\ \ref{fig:198mobdegrade}. After the first measurement,
the device was stored in the nitrogen glove box for about 3 months.
The peak electron mobility decreased by about 17$\%$ at the second
cooldown, which degraded faster than that of sample 194. After the
same NHMFL experiment, the device was in air for about one week, and
the peak electron mobility decreased by another 18$\%$ at the fourth
cooldown. The hole mobility was decreased by about 26$\%$ between
the first and fourth cooldown, measured at 4.2 K [Fig.\
\ref{fig:198mobdegrade}(b)].

After $\sim$ 200 days, the peak electron mobility decreased by about
35$\%$ for both devices. After one week exposure in air, the peak
electron mobility decreased by about 35$\%$ and 18$\%$ for sample
194 and 198 respectively. Roughly it can be estimated that the peak
electron mobility decreased by $\sim$ 0.2$\%$/day in vacuum or a
nitrogen gas environment, and by $\sim$ 5$\%$/day in air.

\begin{figure}
 \includegraphics{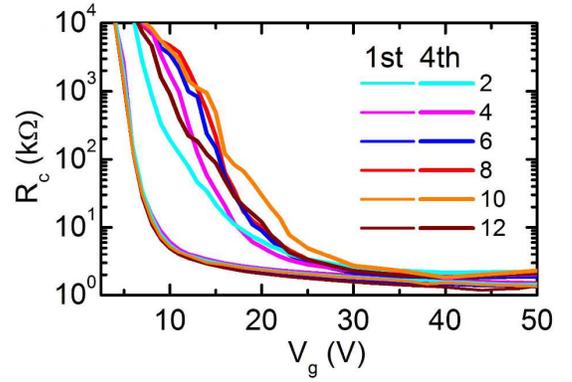}
 \caption{\label{fig:Ncontactdegrade} N$^+$ contact resistance $R_c$
as a function of gate voltage $V_g$ on sample 198 at 300 mK. For 4th
cooldown, at $V_g<15$ V, $R_c$ is calculated when
$|I_{ds}|\leqslant4$ nA; otherwise, $R_c$ is determined at
$|I_{ds}|\leqslant100$ nA. Compared to 1st cooldown, the contact
resistance increased dramatically at low electron density
($5<V_g<25$ V).}
 \end{figure}

In addition to the mobility degradation, the contacts also degrade
with time. The contact degradation makes accurate resistance
measurements difficult, and shows up as the conductance (mobility)
fluctuations both for electrons and holes at lower densities
\cite{KottPhd12, Hu10APL}. In sample 198 (without STI), the contact
problem had been mitigated, but it resurfaced at the 4th cooldown.
The contacts were linear only at above $V_g=16$ V
($n_{2d}=2.5\times10^{11}$ cm$^{-2}$ ) rather than $V_g=6$ V
($n_{2d}=1.02\times10^{11}$ cm$^{-2}$ ) at the first cooldown, and
the contact resistances were also much worse than those at the 1st
cooldown (Fig.\ \ref{fig:Ncontactdegrade}). Correspondingly, the
mobility fluctuated at $n_{2d}<2.5\times10^{11}$ cm$^{-2}$ for the
4th cooldown in Fig.\ \ref{fig:198mobdegrade}(a).

\begin{figure} [!b]
 \includegraphics{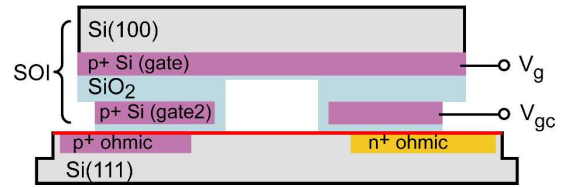}%
 \caption{\label{fig:Bilayergate}A vacuum FET with a bi-layer gate
 structure using SOI. The 2D carrier densities around the contacts
 and at the investigated area can be independently controlled. The
  gate is also fully covered by a SiO$_2$ layer. }
 \end{figure}

Although resistance measurements are possible with pure large
contact resistances ($\sim 10$ M$\Omega$) using 4-terminal DC
method, there probably exist charge trapping and emission processes
in these devices like single electron transistors \cite{KottPhd12}.
The 4-terminal DC method is not effective in this case. In order to
solve the contact problem, the bi-layer gate structure using SOI, as
shown in Fig.\ \ref{fig:Bilayergate} can be adopted. Similar to a
split-gate geometry, the bi-layer gate structure which permits high
densities around the contacts while allowing independent control of
the density of the 2DSs under investigation \cite{[{}][{. The idea
of bi-layer gate structure using SOI is from the discussion with
Alex R. Hamilton of University of New South Wales,
Australia.}]Vitkalov00}, is quite important going forward.

\section{\label{sec:conclusion}PROSPECTS AND CONCLUSIONS}

We have investigated a process to fabricate high mobility ambipolar
H-Si(111) vacuum FETs with electron mobility of $\sim 200,000$
cm$^2$/Vs and hole mobility of $\sim 10,000$ cm$^2$/Vs. From the
electron conductivity vs. electron density analysis, the electron
mobility is still mainly limited by 3D bulk impurities, which
implies that the mobility can be further improved with attention to
high temperature steps. At a minimum, electron mobility of $\sim
300,000$ cm$^2$/Vs should be able to realize in the near future,
since we have already fabricated a device with this mobility using
previous sample fabrication techniques \cite{Kott14PRB}. Combined
with a bi-layer gate structure to solve the contact issue, the
device quality should be comparable to the best SiGe quantum well
devices. In addition, the 2DES has a sixfold valley degeneracy,
which will open up many opportunities for fundamental research and
practical applications.

Three magnetoresistance measurements are a very useful tool to
determine the underlying symmetry of the 2DES. They can be used to
detect possible phase transition when the associate symmetry changes
\cite{Abanin10,Kumar14}. Further experiments with a tilted magnetic
field to re-distribute electrons between valleys can dynamically
change the underlying symmetry. It will surely improve our
understanding of this 2DES and the nature of the observed
anisotropy.

The devices were also investigated at filling factor $\nu<1$ in a
magnetic field up to 35 T. We clearly observed the 2/3 FQH state
with a well developed hall plateau, and a high magnetoresistance
anisotropy at $\nu<2/3$ on both sample 194 and 198. On sample 194,
there were some hints of possible 3/5, 4/7, 5/9 and 6/11 FQH states
(data not shown). With better devices, we should be able to verify
these additional states. It is quite possible to observe new phases
in this unique sixfold valley-degenerate system.

Because of the distinct device structure, the 2D systems are
resident at the surface in the vacuum cavity. Atoms, molecules,
superconductors and other systems can easily couple to the 2DSs and
form hybrid systems \cite{Eng05APL}, and novel functional devices
can be implemented in the future.

\begin{acknowledgments}
This work was funded by the Laboratory for Physical Sciences.
Research was performed in part at the NIST Center for Nanoscale
Science and Technology, and the National High Magnetic Field
Laboratory, which is supported by National Science Foundation
Cooperative Agreement No. DMR-1157490 and the State of Florida. We
also acknowledge the support of the Maryland NanoCenter and its
FabLab. The authors are grateful to Dr. Neil Zimmerman and Dr.
Michael Stewart for their help on the RTA process.
\end{acknowledgments}



%
%

%



%

\end{document}